# Small Satellite Mission Concepts for Space Weather Research and as Pathfinders for Operations

**Amir Caspi[1], M. Barthelemy[2,3], C. D. Bussy-Virat[4], I. J. Cohen[5], C. E. DeForest[1], D. R. Jackson[6], A. Vourlidas[5,7], and T. Nieves-Chinchilla[8]**

[1] Southwest Research Institute, Boulder, CO, USA
[2] CNRS, IPAG, University of Grenoble Alpes, 38000 Grenoble, France
[3] Grenoble INP, CSUG, University of Grenoble Alpes, 38000 Grenoble, France
[4] Department of Climate and Space Sciences and Engineering, University of Michigan, Ann Arbor, MI, USA
[5] The Johns Hopkins University Applied Physics Laboratory, Laurel, MD, USA
[6] Met Office, Exeter, UK
[7] IAASARS, Observatory of Athens, Penteli, Greece
[8] Heliophysics Science Division, NASA Goddard Space Flight Center, Greenbelt, MD, USA

Corresponding author: Amir Caspi (amir@boulder.swri.edu)

**Key Points:**

- Miniaturization of satellite technologies make SmallSats viable, low-cost platforms for space weather research and operational prototyping

- Current missions and proposed concepts show how SmallSats can address relevant space weather measurement requirements

- Suggested paths forward for future implementations using lessons learned from these missions are provided

**Abstract**

Recent advances in miniaturization and commercial availability of critical satellite subsystems and detector technology have made small satellites (SmallSats, including CubeSats) an attractive, low-cost potential solution for space weather research and operational needs. Motivated by the 1st International Workshop on SmallSats for Space Weather Research and Forecasting, held in Washington, DC on 1–4 August 2017, we discuss the need for advanced space weather measurement capabilities, driven by analyses from the World Meteorological Organization (WMO), and how SmallSats can efficiently fill these measurement gaps. We present some current, recent missions and proposed/upcoming mission concepts using SmallSats that enhance space weather research and provide prototyping pathways for future operational applications; how they relate to the WMO requirements; and what challenges remain to be overcome to meet the WMO goals and operational needs in the future. With additional investment from cognizant funding agencies worldwide, SmallSats – including standalone missions and constellations – could significantly enhance space weather research and, eventually, operations, by reducing costs and enabling new measurements not feasible from traditional, large, monolithic missions.

**Plain Language Summary**

Critical technology for satellites and scientific detectors has recently been miniaturized and become commercially available. This has made small satellites (collectively called SmallSats, which includes CubeSats) attractive as low-cost solutions for research into space weather and, potentially, for future forecasting and evaluation of space weather hazards. The 1st International Workshop on SmallSats for Space Weather Research and Forecasting (SSWRF) was held in Washington, DC on 1-4 August 2017. Motivated by this workshop and guided by analyses from the World Meteorological Organization (WMO), we discuss how and why SmallSats can provide advanced measurement capabilities to fill gaps in space weather knowledge. We present some current and upcoming space mission concepts that use SmallSats to make measurements relevant to space weather and provide development pathways for future missions that can fill operational space weather forecasting/monitoring needs. We describe how these missions relate to WMO guidance and what challenges must be overcome to achieve future measurement goals for operational applications. If appropriate technology and infrastructure investments are made by relevant government agencies, SmallSats - including single-satellite missions and multi-satellite constellations - could significantly lower costs and enable new measurements to enhance space weather research and, eventually, forecasting/monitoring operations.

# 1 Introduction

Solar activity drives rapid variations in the radiation and plasma environment in interplanetary and geospace. These variations occur on timescales of minutes and hours, associated with solar flares and coronal mass ejections (CMEs), to a few days, as complex magnetic features on the Sun, such as active regions and coronal holes, rotate across the solar disk. These phenomena result in orders-of-magnitude increases in the fluxes of high-energy (extreme ultraviolet [EUV], and especially X-ray and gamma-ray) photons and energetic, often relativistic particles (electrons, protons, alphas, and heavier ions) streaming through interplanetary space. These enhanced photon and particle fluxes pose direct risks to humans and electronics in space. The increased radiation and associated propagating disturbances in the interplanetary magnetic field (e.g., from CMEs or so-called "co-rotating interaction regions") also drive complex dynamics in Earth's magnetosphere and its ionosphere, thermosphere, and mesosphere (ITM), which pose indirect but significant hazards to aircraft and on-board humans, satellite navigation and orbital stability, radio-frequency communications, and power grids, among other effects. Together, these interrelated processes are often collectively referred to as "space weather," although some terrestrial processes can also affect the near-space in Earth's thermosphere.

Predicting severe space weather events and their effects has, unsurprisingly, become a top priority for numerous government agencies – both military and civilian – and corporate/private institutions worldwide. Recently, the need for space weather readiness has begun to be codified in public policy, e.g., in the National Space Weather Strategy and associated Action Plan documents (NSTC, 2015a,b; Bonadonna et al., 2017) which call for "improving space-weather services through advancing understanding and forecasting" (Goal 5), in particular through "improving forecasting lead-time and accuracy" and "enhancing fundamental understanding of space weather and its drivers to develop and continually improve predictive models" (sub-goals 5.4 and 5.5, respectively).

However, the physical mechanisms underlying space weather phenomena, including the originating drivers at the Sun (e.g., flares, CMEs, and other solar activity) and the resultant dynamical effects induced in Earth's complicated and coupled magnetosphere and ITM, are only partially understood. This significantly limits the accuracy of existing predictive models and subsequent forecasting ability. In large part, progress on this front has been hampered by a lack of measurements with sufficient temporal, spatial, and energy/spectral resolutions and/or sampling, both *in situ* and remote sensing. Improved measurements and systematic studies, and subsequent improvements to forecasting models, are required to improve our understanding of these space weather drivers and effects. This is recognized in the updated National Space Weather Strategy and Action Plan (NSTC, 2019; Knipp & Gannon, 2019), whose Objective II calls for "enhanced, more informative, robust, and cost-effective measurements," including "deploying new and innovative observational platforms and technologies."

Recent advances in miniaturization and commoditization of critical satellite subsystems – including attitude determination and control, high-powered on-board computing, and high-bandwidth communications – and of high-quality detector technology have enabled low-cost, small satellites (SmallSats, including microsats, CubeSats, and other pico-/nanosats) as viable, attractive solutions for targeted measurements to address long-standing scientific questions (Moretto & Robinson, 2008; NASEM, 2016). SmallSats are typically defined as having a mass of $\lesssim$100 kg, although with current ridesharing opportunities enabling masses up to ~300 kg, this threshold is often fuzzy. The defining difference, then, is more philosophical: SmallSats often favor modularity and replaceability through commercially available components. Development and launch costs are thus commensurately lower compared to larger, single, custom-built satellites. This strategy allows measurements not feasible from traditional, expensive platforms – for example, by deploying constellations to enable *in situ* measurements with high spatio-temporal resolution or remote sensing measurements with multiple simultaneous passbands or fields of view; or by rapidly (re)deploying a series of identical craft to provide a continuous measurement record or reduced/real-time data latency. Such capabilities open new windows of opportunity for space weather research and subsequent advances in empirical or physics-based models required for understanding and forecasting space weather phenomena.

The World Meteorological Organization (WMO) specifies measurement requirements for observations of physical variables in support of WMO programs, including for space weather (https://space.oscar.wmo.int/applicationareas/view/space_weather). The requirements are regularly reviewed and updated by a WMO team comprising expert members typically representing their national operational space weather centres. Previously, this team was named the Inter-Programme Team on Space Weather Information, Systems, and Services (IPT-SWeISS), though we note that IPT-SWeISS has since concluded and a new WMO space weather expert team is being formed. The assessments from the expert team routinely indicate that existing observational assets meet the WMO requirements only poorly, and that SmallSat constellations could effectively fill these gaps. These requirements form a framework for advancing technology for space weather understanding and prediction. Current technology and planned space missions do not fully meet these requirements, which therefore may be used to guide current and future space weather initiatives. However, many of the WMO requirements are cost-prohibitive to meet with conventional mission design. Prior CubeSat missions have already demonstrated the feasibility of SmallSats for targeted space weather research (e.g., Spence et al., 2022, and references therein), and hence there is significant interest in investigating SmallSats as a means to satisfy certain WMO requirements at lower cost than, or in ways not achievable by, traditional space missions (e.g., Verkhoglyadova et al., 2021). Bridging WMO measurement gaps and integrating such measurements into a validated and verified forecasting pipeline are necessary steps in establishing SmallSats as viable platforms for space weather operational needs. To that end, the 1$^{st}$ International Workshop on SmallSats for Space Weather Research and Forecasting (SSWRF), held in Washington, DC on 1–4 August 2017, brought together researchers, industry partners, and government agencies to discuss current progress toward, and future pathways for, using SmallSats to improve space weather understanding and prediction.

Here, we discuss several current missions and proposed/upcoming mission concepts, presented at SSWRF, that could bridge gaps in WMO measurement requirements – both for research and prototyping for potential future operational purposes – and that directly address the National Space Weather Strategy and Action Plan's 2019 Objective II. These missions/concepts were largely developed to meet scientific requirements of their own; however, each is relevant to the WMO enterprise and could satisfy corresponding WMO requirements with minor augmentation or could be used as a template for a future low-cost, dedicated, targeted space-weather mission to prototype operational applications of such measurements.

Section 2 provides example WMO requirements and a gap analysis for the existing observational network, using thermospheric measurements as an illustration, followed by discussion of how SmallSats could potentially fill these gaps at low cost compared to conventional space missions using large, monolithic observing platforms. Section 3 details specific current missions and mission concepts from SSWRF, including their objectives, implementation, and how they could address specific WMO requirement gaps. Finally, Section 4 provides a summary and discussion, including the feasibility and maturity of the discussed missions and recommendations to funding agencies for enabling future space weather research and research-to-operations (R2O) pathways with SmallSats.

## 2 Example WMO Requirements and Gap Analysis for Thermosphere Observations

### 2.1 WMO Requirements and Definitions

The WMO space weather requirement list (see URL in §1) includes observations across all categories, including solar, solar wind and particles, geomagnetism, ionosphere, and thermosphere. The requirements imply an emphasis on space weather operations rather than research, and include the observational uncertainty, horizontal and vertical resolution, observing cycle, and timeliness (time to delivery as an operational asset). Each of these categories include three levels of requirement:

- "goal" – the ideal requirement, beyond which no further improvements are necessary;
- "breakthrough" – an intermediate level, which will give significant improvement for targeted applications; and
- "threshold" – the minimum requirement to ensure that the observations are useful.

Along with updating observation requirements, the WMO expert team produces a "Statement of Guidance" (also regularly updated) assessing how adequately existing observations fulfil the requirements and suggesting improvements in space- and ground-based observing systems to fill any gaps. We shall use the thermosphere as an example to illustrate observation requirements and the associated gap analysis. Observations of temperature, atmospheric density, and horizontal wind are required throughout the thermosphere to produce space weather alerts (e.g., of satellite drag). It is also increasingly recognized that a well observed thermosphere can contribute to improved ionosphere forecasts (e.g., Chartier et al., 2013). The current WMO requirements for the thermosphere are detailed in Table 1. The requirements are included separately for "High Thermosphere" (200 to ~600 km) and "Low Thermosphere" (100–200 km) because temporal variations are more rapid and vertical gradients are stronger in the latter region.

| Variable | Layer | Uncertainty | Horizontal resolution | Vertical resolution | Observing Cycle | Timeliness |
|---|---|---|---|---|---|---|
| T | Hi Therm | 35/75/140 K | 100/200/500 km | 20/30/50 km | 5 s / 5 min / 30 min | 30/45/60 min |
| | Lo Therm | 10/14/20 K | 100/200/500 km | 5/10/25 km | 5 s / 60 s / 5 min | 5/20/60 min |
| Density | Hi Therm | 10/15/20 % | 100/200/500 km | 20/50/100 km | 5 s / 5 min / 30 min | 30/45/60 min |
| | Lo Therm | 5/7/10 % | 100/200/500 km | 5/10/25 km | 5 s / 60 s/ 5 min | 5/20/60 min |
| u | Hi Therm | 10/20/30 m s$^{-1}$ | 100/200/500 km | 20/30/50 km | 5 s / 5 min / 30 min | 30/45/60 min |
| | Lo Therm | 5/7/10 m s$^{-1}$ | 100/200/500 km | 5/10/25 km | 5 s / 60 s / 5 min | 5/30/60 min |

**Table 1:** *WMO observational requirements for temperature (T), neutral density and horizontal wind (u) between 200 and ~600 km altitude ("Hi Therm") and between 100 and 200 km altitude ("Lo Therm"). Goal, breakthrough, and threshold requirements are shown (these are the smallest, middle, and largest values, respectively). "Timeliness" is the time for data products to be processed and delivered as operational assets.*

| Variable | Assessment | Comments |
|---|---|---|
| Hi Therm T | Poor | Only a few sparse Fabry-Perot Interferometer (FPI) observations are available. Poor timeliness. |
| Lo Therm T | Marginal | Optical Spectrograph and InfraRed Imaging System (OSIRIS) data are available, but they do not cover whole vertical range and have poor timeliness. |
| Hi Therm density | Marginal | Swarm meets most requirements, apart from timeliness and vertical resolution. Special Sensor Ultraviolet Spectrographic Imager (SSUSI) and Special Sensor Ultraviolet Limb Imager (SSULI) may meet requirements, but no information is available on accuracy, observational cycle and timeliness. |
| Lo Therm density | Less than Marginal / Marginal | SSUSI and SSULI may meet requirements, but no information is available on accuracy, observational cycle and timeliness. |
| Hi Therm u | Poor | Only a few sparse FPI observations. Poor timeliness. Accelerometer winds have too large errors to be useful. Region partially covered by new Ionospheric Connection Explorer (ICON) observations. |
| Lo Therm u | Poor | Data gap (daytime) addressed by ICON. No other current observations. |

**Table 2:** *Gap analysis for temperature (T), neutral density, and horizontal wind (u) between 200 and 600 km ("Hi Therm"), and between 100 and 200 km ("Lo Therm").*

The gap assessment uses the following criteria:

- **Poor** – minimum observing requirements not met, no or limited quality observations provided only by scientific instruments without plans for continuity;

- **Marginal** – minimum requirements met, can be provided by research instruments with existing plans to convert them to operational; and

- **Acceptable** – better than minimum user requirements but less than optimum; operational quality data, with identified risk of discontinuity in data flow

The gap associated with the thermospheric variables is summarized in **Error! Reference source not found.**. It is clear that the thermosphere is currently poorly observed by WMO standards.

## 2.2 Filling the Gap with SmallSat Constellations

The analysis in the previous section indicates the lack of observations of the thermosphere. Satellite observations may provide reasonable horizontal coverage but the vertical range of the observations is limited, while ground-based observations are available at only a few locations. Recent missions such as the Ionospheric Connection Explorer (ICON; Immel et al., 2018) and Global-scale Observations of the Limb and Disk (GOLD; Eastes et al., 2017) help to address some of these issues, but the benefit of these observations for operational applications is likely to be limited: first, in common with many of the other observation types reviewed here, the timeliness of the observations is quite poor; and second, these are one-off research missions which are not ideal when it comes to the long-term development and maintenance of the operational observation network. A further issue is that a limited number of observation systems restrict the capability for independent verification. This has been highlighted in a recent study by Aruliah et al. (2017), who indicated possible biases in accelerometer-based densities compared to density inferred from Fabry-Perot interferometers (FPIs).

In this context, there is a great opportunity for these shortcomings to be addressed via a constellation of SmallSats. An individual SmallSat may not give good vertical coverage of the thermosphere, but, because of their low cost, we can envisage a constellation of SmallSats that together cover a wide range of altitudes. Here, we use the example of the recent QB50 mission (Thoemel et al., 2014; Masutti et al., 2018) to show how this could be improved upon. QB50's objective is to carry out atmospheric research within the lower thermosphere, 200–380 km altitude, by providing multi-point, in-situ measurements for many months. QB50 comprises numerous CubeSats, each flying a range of instruments including an Ion-Neutral Mass Spectrometer (INMS; Bedington et al., 2014), which observes temperature and neutral density.

How well do the INMS observations of neutral density and temperature meet the WMO requirements? The QB50 constellation used Nanoracks to launch 28 CubeSats in 2 batches, 60 days apart, at an altitude of 415 km. This led to a separation in altitude of the order of tens of km (Masutti et al., 2018) which improved the vertical spacing slightly. QB50 also launched 8 CubeSats on the PSLV rocket to an altitude of 500 km, which also helped to improve the vertical coverage. This is a very good example of a mission design team using the WMO **vertical resolution** observation requirements to improve the design and functionality of their system.

The accuracy of the INMS observations is still being assessed. The **horizontal resolution** and **observation cycle** likely meet WMO requirements. The **timeliness** of reception of the QB50 observations is poor, because of the lack of available funds for the ground station network; however, with a growing number of low-cost commercial ground-communication providers entering the market, and/or with a targeted (if expensive) infrastructure investment by relevant agencies, this problem could be overcome. Clearly, the lessons learned from QB50 will be very important in the design of a future operations-focused SmallSat constellation.

The lifetime of the QB50 CubeSats presents another issue. Their orbital altitudes drop in time due to drag, leading to the instruments ceasing to function around 200 km as the spacecraft disintegrate during re-entry. Without propulsion, the lifetime of satellites deployed into the lower altitudes (below ~420 km) will not exceed ~18 months, and likely less, depending on the level of solar activity (a direct driver of thermospheric density, and hence drag). Masutti et al. (2018) showed that 3 of the QB50 CubeSats had already de-orbited less than 1 year after launch. Operational weather satellites are designed for longer lifetimes than this and have backup satellites for redundancy; for example, the EUMETSAT Metop mission launches a new satellite every 6 years (https://www.eumetsat.int/our-satellites/metop-series). Then, to use a constellation of CubeSats such as those used in QB50 in an operational-capable capacity, we need to do either or both of the following. First, we need to have CubeSats flying at as broad a range of altitudes as possible, to ensure that the vertical resolution requirements are met. This may require launching a new constellation every year or so, in order to replenish CubeSats that have de-orbited. Second, there is a need to invest in new small-scale technology, in particular for propulsion/station-keeping, to ensure that the CubeSats will remain within an orbital altitude range for longer and reduce the need for replenishment of the constellations.

Recently, there has been a large rise in commercial multi-satellite constellations to supply broadband internet, including SpaceX's Starlink, and OneWeb. These offer an alternative or additional approach to a targeted constellation for atmospheric measurements: adding thermospheric instruments as hosted payloads on these commerical satellites. Because the satellite operators have specific altitude requirements (e.g., ~550 km for Starlink, ~1200 km for OneWeb), this would not fully meet WMO vertical resolution requirements, but could help fill gaps and reduce deployment costs. Use of complementary technology is another potential gap-filling solution; for example, the proposed Skimsat satellites (Bacon & Olivier, 2017) will fly at ultra-low orbits (around 160 km) and have the potential to carry thermospheric instruments.

## 3 SSWRF Missions to Address Space Weather Research & Operational Prototyping

The WMO requirements provide guidelines to help inform designs of SmallSat missions targeted at space weather research or operations. Even missions intended for other applications, such as solar or terrestrial research, could provide data useful for investigating space weather, and thus can also be considered in the context of the recommended WMO requirements. Here, we discuss several current missions and developed mission concepts that were presented at SSWRF and can be used to address open scientific questions in space weather research and gaps in space weather operations. These missions represent only a small sampling of the many current and future pathfinding SmallSats relevant to space weather (e.g., ELFIN: Angelopoulos et al., 2020; AEPEX: Marshall et al., 2020; REAL: Millan et al., 2020; GTOSat: Blum et al., 2021; SunCET: Mason et al., 2021; and many others) but highlight a broad cross section of WMO-relevant measurements. The missions are presented in rough order of concept age/maturity and, for each mission, we discuss both the current implementation and the challenges and/or technology needs for addressing WMO requirements or for prototyping future operational applications.

### 3.1 Cyclone Global Navigation Satellite System (CYGNSS)

*Mission Objectives* – The CYGNSS constellation was launched on December 15, 2016 (Ruf et al., 2013). Although the primary objective of the mission is to measure surface winds in tropical cyclones, the constellation configuration and the techniques to control the trajectories of the satellites have been used for space weather purposes, namely to investigate thermospheric density at ~500 km.

*Mission Implementation* – The 8 small satellites of the CYGNSS constellation were deployed from a single deployment module. To maximize the coverage of tropical cyclones, the observatories need to be evenly spaced out along the orbit. Since the spacecraft do not include thrusters, passive techniques have been implemented to control their trajectories and position them at equal distances along the orbit. These maneuvers, called differential drag maneuvers, consist of pitching the satellites by ~78° to increase the cross-sectional area with respect to the velocity vector. By doing this, the drag force acting on the maneuvered satellite is multiplied by a factor ~6, altering the in-track velocity, and thus its spacing from the neighboring satellite.

The technique of controlling the small satellite trajectories using atmospheric drag can be used to improve the accuracy of algorithms in determining the thermospheric density from the satellite ephemerides. Each spacecraft includes a GPS receiver which gives its position and velocity at a cadence of 1 s. By applying filtering techniques to these ephemerides, the drag acceleration and atmospheric density can be inferred. The accuracy of this method is greatly improved by comparing the ephemerides of satellites in high drag to those in nominal configuration, for which the drag force is minimal. In the final configuration, the CYGNSS satellites will be evenly spaced out along the orbit, with ~12 minutes separating each observatory (Bussy-Virat et al., 2018). This configuration will allow the detection of small-scale features and short temporal variations of the thermospheric density, which can be of particular importance during, and shortly after, geomagnetic storms.

The benefit of the CYGNSS SmallSat mission design is that it exploits an existing high-precision radio signal to actively probe important elements of the global terrestrial environment, from multiple vantage points, at low cost.

*Mission Challenges and/or Technology Needs* – Determining the atmospheric density by applying filtering techniques on small satellite ephemerides doesn't require a particular type of technology, although the accuracy of the results greatly relies on the precision of the orbit determination process, i.e., on the measured positions and velocities.

There are a few challenges with this technique, though. In the case of the CYGNSS mission, the satellites fly at 500 km, which means that the atmospheric density, and hence also the drag force, is much smaller than at lower altitudes. In addition, solar activity has been extremely low since launch, due to the timing of the mission close to solar-cycle minimum. As a result, the density has been lower than usual. These two effects combined imply that the perturbations of the satellite trajectories due to drag have been considerably small, degrading the accuracy of the filtering techniques in estimating the atmospheric density.

### 3.2 Polarimeter to UNify the Corona and Heliosphere (PUNCH)

*Mission Objectives* – PUNCH is a LEO remote-sensing mission under development through NASA's Explorers Program, with commencement of science operations scheduled for 2024 followed by a 2-year prime mission. PUNCH will comprise a constellation of four SmallSats to produce polarimetric images of the solar corona and inner heliosphere. The goal of the mission is to determine the cross-scale processes that unify the corona and heliosphere. PUNCH collects 3D images (e.g., DeForest et al., 2013; Howard et al., 2013) of the entire inner heliosphere every few minutes over a period of years; these data are immediately applicable to predictions of both arrival time and geoeffectiveness of CMEs. PUNCH exceeds the WMO requirements for heliospheric imaging sensitivity, and can, in principle, meet "timeliness" (latency) requirements developed independently (DeForest et al., 2016); as of Phase C, the mission has been tasked (and funded) to ensure the ground system is able to support that capability, if needed.

PUNCH data could improve the effectiveness of space weather prediction in two principal ways: (1) by tracking 3D location of CME fronts and stream interaction regions (SIRs) directly, avoiding confusion that is intrinsic to stereoscopic or 2D imaging; and (2) by identifying the chirality of CME flux ropes via 3D imaging of embedded density structures (e.g., DeForest et al., 2017). Chirality is important because it is the "missing link" between readily-measured magnetic polarity at the photosphere and prediction of the leading-edge $B_z$ parameter, a major indicator of geoeffectiveness.

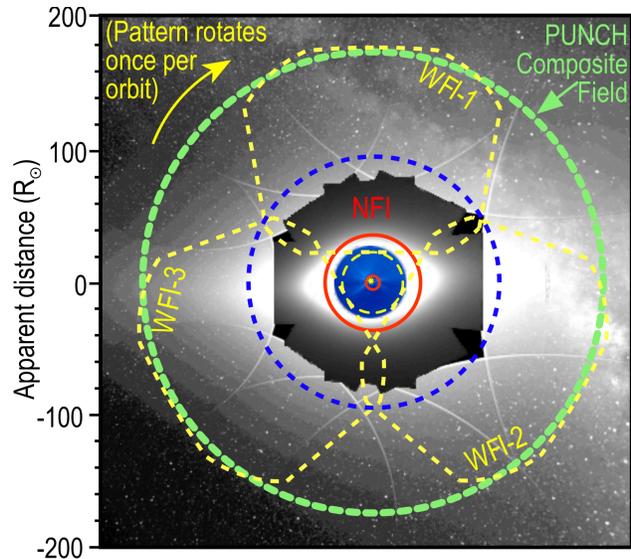

*Fig. 1:* PUNCH observes the corona and inner heliosphere every few minutes with four independent, synchronized, spectrally matched cameras. Images are stitched together by the ground pipeline to produce continuous 3D image sequences of CMEs as they propagate. The imaging data exceed WMO requirements.

*Mission Implementation* – The four PUNCH SmallSats will be deployed from a single launch vehicle in Sun-synchronous dawn/dusk LEO. One of the SmallSats carries a Narrow Field Imager (NFI; 6–32 $R_\odot$) implemented as a coronagraph; the other three each carry a Wide Field Imager (WFI; 18–180 $R_\odot$) with heritage from the STEREO/HI instrument. Data from the four imagers are stitched together digitally on the ground to produce routine 360° round images of the inner heliosphere, from elongations of about 1.5° (6 $R_\odot$) out to 45° (180 $R_\odot$) from the Sun (Fig. 1). The instruments are synchronized to sub-second precision and spectrally matched, to simulate a single "virtual instrument" with a continuous and very broad field of view.

The benefit of the PUNCH SmallSat mission design is that it yields continuous, full-field, high-cadence coverage from LEO, that would otherwise require a deep-space (e.g., L1) mission with far more challenging and expensive environmental and telemetry requirements.

*Mission Challenges and/or Technology Needs* – As proposed, PUNCH's instrumentation meets or exceeds identified space weather R2O and anticipated operational needs. The primary challenge compared to the existing proposed scientific mission is improving the latency of data downlink to the ground. As conceived for NASA, PUNCH ground passes are approximately once daily per spacecraft. This is acceptable for research and R2O-development activities, including retrospective arrival prediction. For demonstration of actual quasi-operational utility, latency of as little as 2–3 hours would be required (DeForest et al., 2016). From a polar orbit, this could be achieved through additional ground passes with no modification to the flight assets. Achieving the WMO latency requirement of 1–2 hours could be accomplished through a future dedicated space-weather mission with additional ground assets.

### 3.3 CubeSat Imaging X-ray Solar Spectrometer (CubIXSS)

*Mission Objectives* – CubIXSS is a 6U Sun-pointed CubeSat (Caspi et al., 2020) selected through NASA's HFORT program in November 2021. Its measurements of the soft X-ray (SXR: ~0.025–5.5 nm; ~0.23–50 keV) solar spectral irradiance have distinct space weather applications (Caspi et al., 2015). In particular, the 1–5 nm wavelength band is highly variable with solar activity (Rodgers et al., 2005) and also drives significant dynamics in the ionosphere D- and E-regions (Sojka et al., 2013, 2014). The specific details of the dynamics depend strongly on the spectral distribution, which to date has never been fully measured well, with prior observations being severely limited in some combination of spectral resolution, passband, cadence, or overall duration; this significantly limits understanding of how ionospheric dynamics are driven by solar SXR forcing, both on minute-to-hour timescales from solar flares and on few-day timescales from active region evolution and solar rotation. CubIXSS measurements will fill this observational gap.

*Mission Implementation* – CubIXSS comprises a suite of instruments (Fig. 2) packaged into a 6U CubeSat to be flown in LEO for a ~1-year prime mission. Four miniature X-ray detectors with heritage from the Miniature X-ray Solar Spectrometer (MinXSS) CubeSats (Mason et al., 2016, 2020; Woods et al., 2017; Moore et al., 2018) measure full-Sun SXR spectral irradiance over a combined range of ~0.5–50 keV (~0.025–2.5 nm) with a spectral resolving power of 13–88 and cadence down to ~1 s. The apertures and entrance filters of these detectors are tailored to match the full dynamic range of SXR flux from solar-cycle minimum up to intense, X-class flares that are significantly geoeffective. A novel pinhole-camera spectral imager extends solar spectroscopic measurements down to 0.23 keV (up to 55 Å), a range never-before routinely measured, with higher resolving power of 14–136 and cadence down to ~20 s, and can measure spectra from individual active regions. These spectral and temporal resolutions exceed the WMO "goal" requirements.

The benefit of the CubIXSS SmallSat mission design is that it collects important space-weather-relevant measurements from a very low-cost, miniaturized platform at a small fraction of the cost and development time of a conventional mission (Caspi et al., 2016).

*Mission Challenges and/or Technology Needs* – As proposed, CubIXSS can be implemented with current technology to meet space weather research needs. While CubIXSS requires fine pointing and high data rates, these are available from current, on-market, commercial-off-the-shelf (COTS) solutions. However, meeting operational needs – specifically, the "timeliness" (latency) requirement for data availability – poses a significant challenge. Satellite-to-satellite communications technology could provide effectively real-time communication, but existing CubeSat-compatible technologies (e.g., GlobalStar) are severely limited in bitrate and therefore would not allow sufficient cadence to meet WMO requirements. Higher-bandwidth sat-to-sat comms would be required to enable downlink of spectral images at the required cadence (with pipeline processing on the ground), and in a small and affordable package. X-band transmitters exist on-market, and both X-band transceivers and SmallSat-compatible optical communications terminals are currently under development. An alternative option would be to downlink continuously to a broad network of ground stations; this is nearly achievable today using existing low-cost, commercial ground networks such as ATLAS, but such continuous downlink from LEO with appropriate reliability would need to be demonstrated.

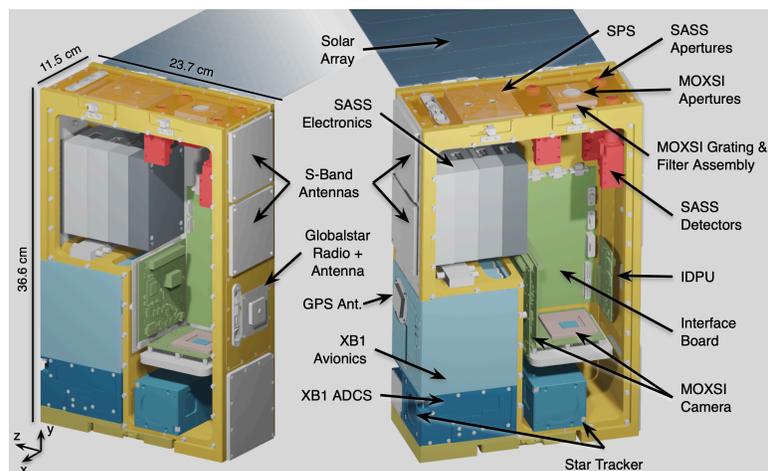

*Fig. 2:* Layout of CubIXSS showing all internal components; the MOXSI pinhole imager uses much of the rightmost 3U.

As proposed, CubIXSS is a single-spacecraft mission. In a sun-synchronous dawn/dusk polar orbit, solar observations would be continuous for most the year, although a few-minute eclipse period would occur in each orbit for a few weeks near vernal equinox for LEO altitudes. In a mid-latitude inclination orbit, significant eclipses (~35% of the orbital period) would occur each orbit. Thus, to meet requirements for continuous coverage, at least two spacecraft would be needed, with orbits sufficiently spaced to ensure that their eclipses do not overlap. Excluding non-recurring development costs, multiple copies could be built for ≲$2.5M each, and thus many spacecraft could be deployed to provide continuous observations over multiple years, for much less than a traditional large mission budget.

### 3.4 Rocket Investigation of Current Closure in the Ionosphere (RICCI)

*Mission Objectives* – RICCI is a suborbital sounding rocket mission concept that will use CubeSats, deployed from the rocket (Cohen et al., 2020), to make multi-point magnetic field measurements to make the first-ever direct measurement of the altitude structure of ionospheric currents via the curlometer technique (e.g., Dunlop et al., 2002) and assess their impact on magnetosphere-ionosphere coupling, specifically on electromagnetic energy dissipation and Joule heating. RICCI measurements are directly relevant to several WMO ionospheric space weather observations tied to ionospheric density (i.e., "*hmF2*", "*foF2*", "*h'F*") and the targeted ionospheric characteristics are intimately coupled to atmospheric processes as ionosphere/thermosphere heating is sensitive to the altitude structure of the ionospheric currents.

*Mission Implementation* – The RICCI objectives will be accomplished by two rockets targeting a stationary (or near-stationary) east-west nightside auroral arc, providing a simple configuration of the auroral electrodynamic system (Fig. 3). The first RICCI rocket will carry a payload of canisters containing trimethyl aluminum (TMA) gas and four 2U CubeSat miniature sub-payloads to be deployed to form a tetrahedral formation with 3-km (baseline) separations. The multipoint measurements from this tetrahedron will enable the first direct in-situ measurement of the altitude profile of the ionospheric currents associated with an auroral arc via application of the curlometer technique.

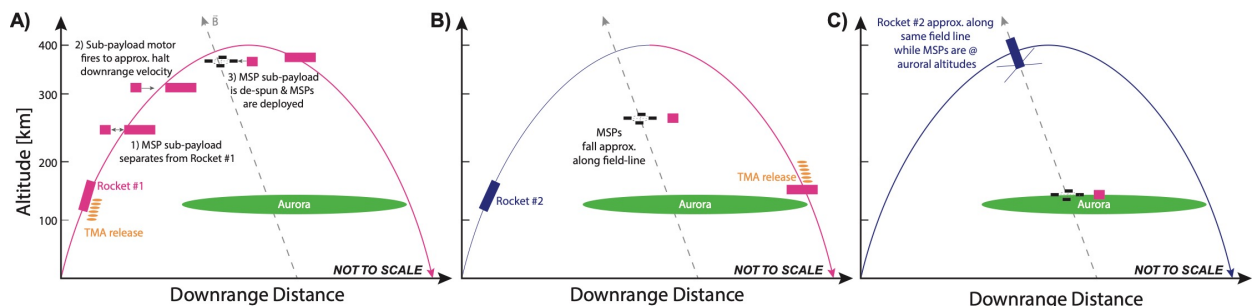

**Fig. 3:** *Schematic diagram of the RICCI mission concept (updated from Cohen et al., 2020).*

The second RICCI rocket will carry an array of vector fluxgate magnetometers to measure the magnetic field, an electric field instrument to measure the two-dimensional electric field, an electron top-hat electrostatic analyzer to measure precipitating auroral electrons, and a new topside ionospheric sounder to measure altitude profiles of the ionospheric electron density. The launch will be supported by simultaneous ground-based imaging and radar observations. Targeting a nominal apogee of 400 km for each rocket, the RICCI mission will enable the first direct measurement of the altitude distribution of the ionospheric closure currents from approximately 200 to 90 km. The launch criteria include a combination of real-time upstream solar wind monitoring and auroral imaging at Poker Flat Rocket Range and downrange (e.g., Fort Yukon or Venetie) to provide additional information on temporal and spatial evolution of the aurora. Additional ground-based PFISR measurements will provide derived height-resolved electric fields as well as context of the large-scale auroral electrodynamics configuration.

The benefit of the RICCI mission design is that formation flight of CubeSats can yield local field-curl and field-gradient measurements inaccessible via other means, at very low cost and complexity compared to conventional constellation missions.

***Mission Challenges and/or Technology Needs*** – Application of the curlometer technique requires that the uncertainty in the field measurement be less than the difference between the measurements at any two points. This is challenging at ionospheric altitudes because of the strength of the background magnetic field, which can give rise to very large uncertainties if attitude knowledge is not known to very high precision (~0.07° per km tetrahedral separation). As such, previous attempts using multi-point *in-situ* ionospheric magnetic field measurements to derive ionospheric currents have been limited by attitude knowledge uncertainty (e.g., Zheng et al., 2003; Martineau et al., 2015). Existing COTS attitude determination systems with sufficient accuracy to meet the attitude knowledge requirement use star trackers that cannot accommodate high rates of payload motion/spin. However, rocket sub-payload technologies currently available at NASA Wallops Flight Facility use high rates of spin for stabilization.

To address this challenge, the RICCI mission will use tailored CubeSats as miniature sub-payloads deployed from the sounding rocket. CubeSats have yet to be deployed from a suborbital sounding rocket platform and adapting them for a sounding rocket poses some challenges, specifically to the operations concept and magnetic cleanliness; however, the COTS subsystems and components available for the CubeSat platform address the technical challenges outlined above and provide the best opportunity to investigate ionospheric closure currents without significant development of sounding rocket sub-payload technologies.

### 3.5 Auroral and airglow monitoring missions

In the range of missions which can fill the gap, we must consider optical surveys of the aurora and airglow. Since auroral optical emissions are mainly due to energetic inputs (photons on the dayside and suprathermal electrons and protons) in the ITM, and are sensitive to eV- and keV-range particles, they are a good way to reconstruct these inputs and their deposition into the upper atmosphere over wide auroral regions. While tools to reconstruct electron fluxes are not yet operational, pathfinding measurements of optical emissions can help to test interpretation models to develop such tools and fill gaps in global electron-input surveys.

With increasing miniaturization, imagery and spectroscopy are now accessible to CubeSats from 2U to 12U. Monitoring aurora and airglow in this manner presents an interesting complement to measurements performable by CYGNSS (§3.1) or CubIXSS (§3.3) in the sense that it gives

Manuscript submitted to *Space Weather Journal*

Caspi et al.  SmallSat Missions for Space Weather Research  13
access remotely to the 90–300 km altitude region, below the region where CYGNSS probes and which includes the altitudes where the solar soft X-rays measured by CubIXSS are preferentially deposited. While there is currently no WMO requirement for auroral monitoring, a real-time auroral imaging or modeling product would provide knowledge of polar cap absorption and Arctic GNSS service impacts during major geomagnetic storms. Real-time imaging products to fulfill forecasting or nowcasting needs would almost certainly require large constellations to enable large-area measurements with low latency, an ideal application of SmallSats. To fill the gap in the meantime, models such as OVATION-Prime do exist, but do not capture the full complexity of true auroral dynamics (Mooney et al., 2021). Single-point pathfinding measurements can provide benchmarks against which such models can be tested and improved.

Two missions are presented as examples for this topic: AMICal Sat and ATISE (Barthelemy al., 2018). Highlights are given below; these two examples show that very sensitive optical instruments can be built for CubeSats, which therefore provide a compelling space weather application, especially for future auroral monitoring, e.g., from constellations of such CubeSats or from hosted payloads leveraging commercial mega-constellations such as Starlink.

### 3.5.1 Auroral and Moon Intensity Calibration Satellite (AMICal Sat)

*Mission Objectives* – AMICal Sat is dedicated to monitoring of the auroral oval. Both nadir and limb observations will be performed. The wide field imager (40°) allows a large view of the auroral oval, giving some constraint of the oval extension and thus on large magnetospheric processes. Despite this wide field, the spatial resolution from LEO is better than 2 km at the green level (120 km), allowing small-scale link to the magnetosphere. AMICal Sat will also allow limb observation with a vertical resolution of 5 km at the limb. This observation geometry is interesting since it gives access to the vertical profile of the emissions and thus to the energy deposition of magnetospheric particles in the ITM. This will mainly scan particles in the eV and keV ranges for electrons and in the keV range for protons and thus also enables measurements of secondary suprathermal electrons.

*Mission Implementation* – AMICal Sat is a 2U CubeSat carrying a sparse RGB (Red/Green/Blue) detector, meaning that only 1 pixel over 16 has a colored filter while the other 15 are black and white. Combined with large pixel pitch (10 µm), this enables a high sensitivity, allowing acquisition of photometrically calibrated auroral images in less than 1 s exposure time, using an objective with a focal length $f = 23$ mm and an aperture of $f/1.4$. AMICal Sat launched into Sun-synchronous orbit on 3 Sep 2020 on Vega flight VV16. Despite failure of the attitude determination and control system (ADCS) after 10 days, it regularly obtained and downlinked images of the auroral oval, and was still operational as of August 2021. First results show that using only an RGB imager represents a loss of information which does not allow reconstruction of energetic inputs with sufficient accuracy on all configurations (Barthelemy et al., 2022). Preserving spectroscopic information, as ATISE (§3.5.2) will do, is important.

*Mission Challenges and/or Technology Needs* – AMICal Sat was designed to be a rapid-development mission with high quality science case and space weather uses. For optimal science, the imager requires an especially short exposure time (1 s) to resolve the timescales of dynamics within the aurora. This exposure time also avoids motion-induced blurring from the coarse pointing control available from a typical ADCS used on 2U CubeSats, primarily controlled by magnetorquers. (High-precision ADCSs are on-market from Blue Canyon Technologies and have flight heritage, e.g., Mason et al., 2017, but these are not currently available outside the U.S. due to export restrictions.) Such a short exposure necessitates an objective with large aperture ($f$/1.4 or, optimally, $f$/0.95) in a very compact design, as well as a detector with large pixels. This necessitates a trade-off between spatial resolution and light-gathering power. The chosen design retains a km-level spatial resolution.

### 3.5.2 Auroral Thermosphere Ionosphere Spectrometer Experiment (ATISE)

*Mission Objectives* – ATISE shares the mission objectives of AMICal Sat but with a focus on spectroscopy. Spectral information is extremely important for auroral monitoring since energetic inputs cannot be fully reconstructed with only RGB information or imagery in a single spectral line. Spectroscopy enables discrimination between the atmospheric response of each element (O, $N_2$, $N_2^+$, NO, $O_2$, etc.) constituting the atmosphere. A short exposure time is required to resolve the auroral dynamics on relevant timescales. ATISE is designed to measure the full visible and near-UV (NUV) spectrum of the aurora at six different altitudes, with exposure times of 1 s, to allow a better reconstruction of the altitude- and time-dependent energetic inputs and energy deposition in the auroral oval.

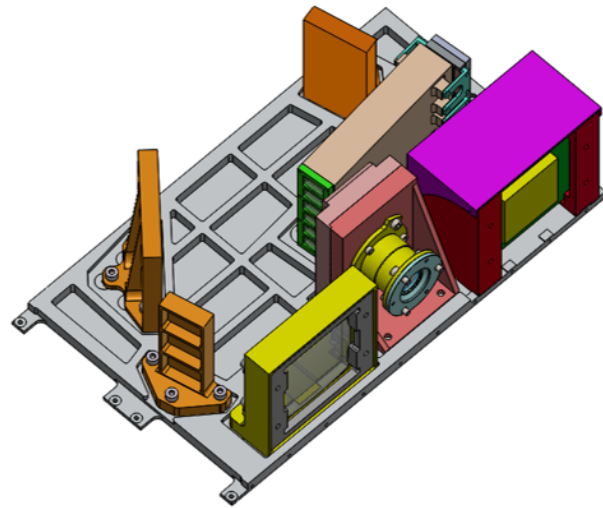

*Mission Implementation* – ATISE is a multi-line-of-sight spectrometer that is extremely sensitive and very compact (Fig. 4). Its primary instrument fits in 6U, and the entire spacecraft is only 12U. The instrument is a Fourier Transform spectrometer based on a Fizeau interferometer using the µSPOC (Micro Spectrometer on a Chip) concept (Diard et al., 2016). In its current version, ATISE has 6 lines of sight (LOSs) distributed along a vertical, each with a field of view of 1° in the vertical direction and 1.5° in the horizontal direction. This corresponds to an extension of 45 km at the limb. The main strength of the ATISE instrument is its sensitivity. It requires only 1 s of exposure time to measure a full auroral visible spectrum (370–900 nm) with a spectral resolution of ~1 nm and a detection threshold of 5 R. Another advantage is its very wide dynamic range (>$10^6$). The central part of ATISE comprises a block of 3 detectors (3 MPix each) with 2 Fizeau interferometers on each detector. This therefore requires some on-board data reduction, achievable with now-standard CubeSat on-board computers (OBCs).

*Fig. 4: Mechanical view of ATISE. The grey plate size is 20×30 cm. The height is less than 10 cm. The spectrometer takes 4U, the imager 1U, and the control electronics 1U.*

To improve the science return, an imager (e.g., like the one in AMICal Sat) will be flown alongside the spectrometer on the 12U CubeSat. This provides context imagery to see where the spectrometer is aimed and to help determine the type of auroral structure from where the spectra are being emitted. ATISE is planned for launch in 2023, and will help to validate and improve electron-deposition models of auroral optical emissions for future operational applications.

*Mission Challenges and/or Technology Needs* – ATISE requires a stable and high-precision ADCS, as well as high downlink availability with low latency to enable monitoring at or near real-time. An orbital inclination of ~70° would be preferable compared to a sun-synchronous orbit, as this would enable measurements of conditions at every local hour rather than a fixed time.

Technological challenges include requiring very low noise on the detectors to enable measurements of weak signal, as well as tight thermal control on the spectrometer since the measurement requirements allow only a very small temperature gradient in the central part of the instrument. Implementing this instrument to meet all of the above requirements within a relatively small (6U) volume, with sufficient reliability to operate in space for the required lifetime, also presents a challenge.

### 3.6 Interplanetary SmallSat Constellations

Here we discuss how small satellites could be used in space weather prediction and forecasting, rather than nowcasting. To be effective, small satellites must be able to monitor the sources of solar activity and the resulting transients in the inner heliosphere. The primary measurement requirement for space weather forecasting is imaging of CMEs close to the Sun, followed by improved characterization of the photospheric magnetic field over the full solar surface, which is the primary boundary condition for heliospheric models. Both types of measurements are optimal when obtained from viewpoints away from the Sun-Earth line or the ecliptic plane (e.g., Vourlidas, 2015; Gibson et al., 2018). These observations are currently performed by imaging and spectroscopic instruments mounted on standard size spacecraft and (mostly) deployed in deep space. To date, this has not been a regime where small satellites could operate, and no solar imager or visible-light/UV spectrometer has yet been miniaturized and flown in space. But large missions are few and far between, and improvements in space weather forecasting generally require increased coverage (remote and in-situ) throughout the inner heliosphere. Solar observing is a challenging area for small satellites but it can bring great benefits to space weather operations, as we detail with a few example mission concepts below.

#### 3.6.1 Fractionated Space Weather Base at Sun-Earth Lagrangian L5 point

*Mission Objectives* – This mission is designed to provide early detection of Earth-bound CMEs and measure their kinematics below 20 $R_\odot$; to provide a 3-to 4-day advance warning for recurrent disturbances and irradiance variations; and to improve the modeling of the inner heliospheric solar wind and magnetic field structure. It uses a modular swarm-of-SmallSats approach to avoid drawbacks of monolithic space probes. The Fractionated Space Weather Base is derived from a monolithic satellite concept studied in the 2013–2022 Heliophysics Decadal Report (NRC, 2013), and adopts that concept's objectives related to space weather research. These objectives can be adjusted and expanded thanks to the unique adaptability of the fractionated mission concept.

*Mission Implementation* – The fractionated mission concept distributes the major components of a standard monolithic spacecraft into several smaller satellites. A strawman concept involving solar sails and a constellation of five 6U CubeSats has been studied by Liewer et al. (2014). The constellation consists of: (1) the communications hub which carries a high-gain antenna and hardware necessary to collect data from the other four science members; (2) a white-light telescope (heliospheric imager) to image CMEs; (3) a full-disk line-of-sight magnetograph to measure the photospheric magnetic field; (4) a solar wind plasma instrument and magnetometer to measure the local solar wind; and (5) an energetic particle instrument to measure solar energetic particle populations. The first three spacecraft are 3-axis stabilized and the in-situ ones are spin stabilized (after arriving at L5). Each CubeSat weights ~10 kg and allocates ~2U for the solar sail, ~2U for the common subsystems (avionics, attitude control, etc.) and ~2U for the science payload. The constellation can be launched towards L5 individually or in groups depending on launch availability, and can cruise to station with current-technology solar sails (e.g., Lightsail-1 and -2). The cruise phase lasts about 3 years. The constellation members orbit in loose formation (~1000 km), which is readily achievable and requires minimal station-keeping around the L5 point (Lo et al., 2010).

The fractioned concept presents many advantages over monolithic missions: (1) it allows for straightforward replacement of failing members and upgrades with better or newer instruments as technology evolves (for example, the 6U-compatible coronagraph presented by Korendyke et al., 2015); (2) it considerably reduces spacecraft and instrument requirements, such as magnetic cleanliness or pointing, since in-situ measurements benefit from spinning and imaging observations prefer stable pointing; (3) integration and testing (I&T) becomes simpler and faster as it can be parallelized and performed in different institutions; (4) schedule pressure is reduced since the different members can be launched at different times; (5) international cooperation (and associated savings) is much easier since each country can build its own payload; and (6) the constellation can serve as a prototype for ingesting L5 (and L4) observations into the forecasting workflows and for prototyping a permanent space weather base at L5/L4.

A variation of the fractionated concept is the 'flock' concept wherein a mothership containing the communication systems and non-miniaturized telescopes carries CubeSats to station and deploys them in loose formation around the mothership itself.

The benefit of a fractionated SmallSat mission design is that it maintains the heritage, maturity, and resources of monolithic systems and payloads, while reducing I&T, schedule, and development costs by reducing engineering constraints on each individual system, including potentially reducing the need for complex subsystems such as propulsion within individual constellation members; by exploiting volume efficiencies in spacecraft production; and, in principle, by allowing ongoing piecemeal replacement of an observatory 'flock.'

***Mission Challenges and/or Technology Needs*** – The two main challenges are the short lifetime of CubeSats and the generally limited radiation tolerance of their subsystems. However, two interplanetary CubeSats launched aboard the Insight mission to Mars in 2018, survived the 6-month trip through the harsh environment of interplanetary space, and functioned successfully until the last contact at the end of 2018. These twin Mars Cube One (MarCO; Baker et al., 2019) CubeSats demonstrated deep-space communication with the deployment of an X-band antenna, and propulsion as they navigated towards Mars on their own. MarCO were based on the same design used in the Liewer et al. (2014) fractionated concept and thus have demonstrated that a CubeSat mission to interplanetary space is viable. However, the L5 members will need to survive for at least 5 years, something that has not yet been demonstrated. Solar sail deployment with the required attitude control system needs to be demonstrated in space as well. Additional technology development is needed to ensure inter-spacecraft communications. A twin spacecraft demonstration mission with the communications hub and a science CubeSat into interplanetary space would probably suffice to demonstrate all key systems and requirements for the L5 mission (solar sail deployment and navigation, inter-spacecraft communication, etc.) while simultaneously serving as the seed for developing a validated forecasting workflow for operational needs, thus opening the path for truly operational missions to the Lagrange L4/L5 points.

### 3.6.2　Small-Scale Structure of Transients (S$^3$T)

***Mission Objectives*** – The mission concept aims to understand (1) the fine-scale structure of transients and (2) the solar energetic particle (SEP) longitudinal distribution at 1 AU. Multipoint in-situ measurements of CMEs have shown that the internal structure of these transients is very complex (Lugaz et al., 2018) with spacecraft separated by as little as 0.01 AU encountering very different magnetic structures. Similarly, SEP measurements from the widely distributed STEREO spacecraft have revealed that SEPs have surprisingly wide longitudinal spreads (Lario et al., 2018), but we know very little about their variation on smaller angular scales. Both of these issues can be addressed by increasing the number of in-situ sensors along the path of incoming CMEs or SEPs.

*Mission Implementation* – The strawman concept calls for a constellation of four spinning 6U CubeSats, initially comprising two SEP and two plasma and fields (P&F) packages. The CubeSats are released into 1 AU orbits with small drifts (~2°/year) relative to Earth. Optimally, the constellation will be equally distributed ahead and behind Earth (Fig. 5). For optimal coverage, the SEP CubeSats are deployed first because the P&F measurements require smaller angular separations than the energetic particle measurements. There is no need for propulsion. The constellation can be augmented with continual launches of the same payloads to maintain a dense longitudinal coverage.

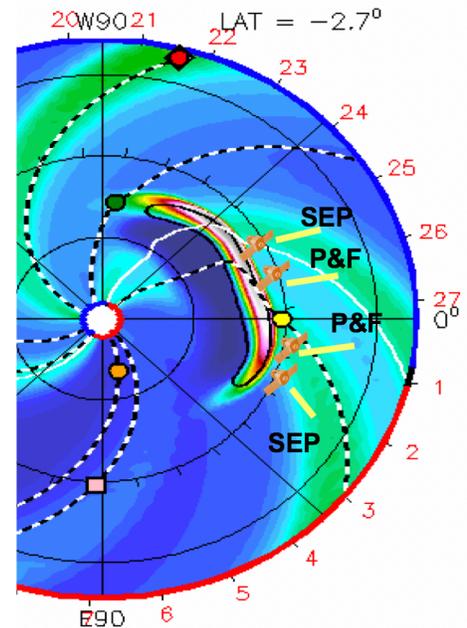

*Fig. 5: A sketch of the S³T concept showing the possible distribution of a constellation of four spinning CubeSats at 1 AU. Two carry SEP detectors, while two carry P&F instruments. The background is an ENLIL model showing a transient impinging on Earth (yellow dot).*

*Mission Challenges and/or Technology Needs* – The concept requires development and/or demonstration of the same technologies discussed in the previous section, namely, interplanetary CubeSats (lifetime, radiation tolerance) and inter-spacecraft communications. If designed with an operational mindset, i.e., with low data latency and regular constellation replacements, S³T could form the basis for augmenting forecasting workflows with multipoint in-situ measurements, something that it is not possible presently.

## 4 Discussion and Conclusions

SmallSats are poised to launch a potential revolution in space-borne scientific endeavors, particularly for space weather research and, eventually, operations that would benefit greatly from the multi-point measurements, low cost, and rapid replaceability offered by SmallSat platforms. By lowering barriers to mission development through lower part, labor, and launch costs, they offer both new, formerly infeasible capability, and new types of mission flexibility. Coupled with rapid advancements in enabling technology and significant projected cost decreases driven by the burgeoning commercial space services industry, it is likely that some operational goals of space weather R2O with SmallSats could be achieved within a decade or less.

We have discussed several current missions and concepts in varying states of maturity, which were presented at SSWRF and are relevant to advancing space weather understanding, monitoring, operational mission prototyping, and/or future predictive capability. Each of the presented mission concepts would advance space weather science in identified ways, either satisfying or making progress toward the WMO-identified observing requirements for improved space weather operations and forecasting. Further, these missions do or will demonstrate new capabilities that are not only feasible, but also natural, to implement with SmallSats, and in particular that are either not pragmatic or not possible with traditional monolithic space missions. Several of the presented missions have objectives specifically targeted towards answering particular space weather-related science questions; these missions can be repurposed and/or slightly augmented to achieve certain WMO requirements (and therefore advance understanding of space weather) at a small fraction of the cost of a larger, dedicated mission.

The largest single benefit of the SmallSat approach to space missions is reduced cost – from development to launch, and potentially to operations – which in turn enables new types of flexible mission and capability development that would be too costly to realistically implement via a traditional high-reliability, low-risk, centralized or monolithic mission design. Some concepts (e.g., CubIXSS, AMICal Sat, ATISE) benefit from reduced cost of a single observatory or instrument that implements a centralized mission at lower cost than traditional larger missions. Many copies could be built and launched – perhaps even with iteratively-improved designs – at a rapid pace to provide continuous temporal coverage and/or reduced data latency. Other concepts (e.g., CYGNSS, PUNCH, RICCI, and $S^3T$) leverage the reduced cost profile of SmallSats to implement constellations to perform coordinated measurements that could not otherwise be feasibly carried out. The fractionated L5 mission concept further reduces cost by rigorously isolating the engineering of independent instruments within an L5 observatory, and by enabling piecemeal replacement of instruments within a single observing 'flock.' This piecemeal replacement is a particularly powerful concept that could yield operational-class reliability from a collection of redundant, individually less reliable, component SmallSats. It is applicable both to high-capability flocks and to individual standalone missions, which would realize reliability benefits from "spacecraft-level redundancy" and on-the-fly replacement, exploiting the ability to improve and replace the entire platform with fast turnaround and at low cost in lieu of the more expensive and complex internal redundancy of key spacecraft systems, required by current strategies for risk reduction in monolithic operational designs.

The scientific successes of a number of recent CubeSat missions (Spence et al., 2022) and the maturity of several of the concepts presented here indicates that SmallSats can legitimately support space weather science and R2O objectives. Pointing stability, on-board computing power, power generation, and component reliability for SmallSats have progressed to the point where these are no longer significant risks or bottlenecks. In some cases, the presented mission concepts could demonstrate theoretical operational capabilities with only minor modifications. However, additional technological development is still required for a number of others, in particular, to increase data downlink capacity, increase data "timeliness" (reduce downlink latency), and enable reliable intra-constellation (satellite-to-satellite) communications. High-speed RF communications solutions for SmallSats in S- and X-band already exist, but these still rely on visibility of ground stations and, due to the limited gain possible within the small footprint available to an on-board SmallSat antenna, are of limited utility for intra-constellation use. LEO-to-MEO or LEO-to-GEO solutions could enable real-time communication without concern for ground-station visibility, but CubeSat-compatible solutions are still rare and currently limited to very restricted data rates. Optical (laser) communication terminals are being developed for SmallSats and present an enticing solution for low-power, high-bandwidth satellite-to-ground communications, including from deep space (e.g., the fractionated L5 or $S^3T$ concepts), as well as for intra-constellation communications.

Funding agencies such as NASA and ESA are beginning to recognize the utility and potential of SmallSats for space weather research; operational agencies such as NOAA also stand to benefit (for example, NOAA has considered SmallSats in its Satellite Observing System Architecture Study; https://www.space.commerce.gov/business-with-noaa/future-noaa-satellite-architecture/). Continued investment in development and maturation of enabling technologies and processes – such as high-speed communications, miniaturization, and multi-unit builds – by all relevant agencies is strongly recommended, as is development of international plans to ease collaboration and coordination of space weather research and operations (Nieves-Chinchilla et al., 2020; Verkhoglyadova et al., 2021). In the near term, targeted space weather operational needs (monitoring and forecasting of specific space weather phenomena) could be partially met by leveraging science-oriented missions such as the ones described above. Data from such missions could be used to enhance forecasting models and/or to test and validate data ingestion into forecasting pipelines. These initial missions can then serve as testbeds for dedicated, operationally-oriented SmallSat missions and associated infrastructure that enable new space weather measurements at lower cost than, with more flexibility than, and potentially in parameter space not explorable by, the current traditional operational strategy of large, monolithic, redundant platforms. With proper direction from and investment by the cognizant funding agencies, the next few years could bring significant new capabilities online.

**Acknowledgments**

A.C. was partially supported by NASA grants NNX14AH54G, NNX15AQ68G, NNX17AI71G, and 80NSSC19K0287. M.B. was partially supported by CNES and by the Programme National PNST of CNRS/INSU, co-funded by CNES. C.B.V. was supported by NASA grant NNL13AQ00C. I.C. was funded through internal investment by the APL Space Exploration Sector. A.V. was supported by NASA grant 80NSSC19K1261. We thank the organizers of the 1$^{st}$ International Workshop on SmallSats for Space Weather Research and Forecasting (SSWRF), and NSF award 1712718 for workshop funding support. No data were used or generated in preparing this manuscript.

**References**

Angelopoulos, V., Tsai, E., Bingley, L., Shaffer, C., Turner, D. L., Runov, A., et al. (2020). The ELFIN Mission. *Space Science Reviews*, 216, 103. https://doi.org/10.1007/s11214-020-00721-7

Aruliah, A. L., Foerster, M., Doornbos, E., Hood, R., & Johnson, D. (2017). *Comparing High-Latitude Thermospheric Winds From FPI and CHAMP Accelerometer Measurements*. Paper presented at AGU Fall Meeting 2017, Abstract #SA41A-2613, New Orleans, LA.

Bacon, A., & Olivier, B. (2017). Skimsats: bringing down the cost of Earth Observation. In: S. Hatton, (Ed.), *Proceedings of the 12th Reinventing Space Conference* (pp 1–7). Cham, CH: Springer Cham. https://doi.org/10.1007/978-3-319-34024-1_1

Baker, J., Colley, C. N., Essmiller, J. C., Klesh, A. T., Krajewski, J. A., & Sternberg, D. C. (2019). MarCO: The First Interplanetary CubeSats. Paper presented at EPSC-DPS Joint Meeting 2019, Abstract #EPSC-DPS2019-2009.

Barthelemy, M., Kalegaev, V., Vialatte, A., Le Coarer, E., Kerstel, E., Basaev, A., et al. (2018). AMICal Sat and ATISE: two space missions for auroral monitoring. *Journal of Space Weather and Space Climate*, 8, A44. https://doi.org/10.1051/swsc/2018035

Barthelemy, M., Robert, E., Kalegaev, V., Grennerat, V., Sequies, T., Bourdarot, G., et al. (2022), AMICal Sat: A sparse RGB imager on board a 2U cubesat to study the aurora. *IEEE Journal of Miniaturization for Air and Space Systems*, submitted (revised). https://arxiv.org/abs/2201.06973

Bedington, R., Kataria, D. O., & Smith, A. (2014). A highly miniaturized electron and ion energy spectrometer prototype for the rapid analysis of space plasmas. *Review of Scientific Instruments*, 85(2), 023305. https://doi.org/10.1063/1.4865842

Blum, L., Kanekal, S., Espley, J., Jaynes, A., Gabrielse, C., Sheppard, D., et al. (2021). *GTOSat: A Next-Generation CubeSat to study Earth's Radiation Belts*. Paper presented at 43rd COSPAR Scientific Assembly, Abstract #2274.

Bonadonna, M., Lanzerotti, L., & Stailey, J. (2017). The National Space Weather Program: Two decades of interagency partnership and accomplishments. *Space Weather*, 15(1), 14–25. https://doi.org/10.1002/2016SW001523

Bussy-Virat, C. D., Ruf, C. S., & Ridley, A. J. (2018). Relationship Between Temporal and Spatial Resolution for a Constellation of GNSS-R Satellites. *IEEE Journal of Selected Topics in Applied Earth Observations and Remote Sensing*, 12(1), 16–25. https://doi.org/10.1109/JSTARS.2018.2833426

Caspi, A., Shih, A. Y., Warren, H. P., Stęślicki, M., & Sylwester, J. (2016). Diagnosing Coronal Heating Processes with Spectrally Resolved Soft X-ray Measurements. White paper submitted to the *Scientific Objectives Team of the Next-Generation Solar Physics Mission*. https://arxiv.org/abs/1701.00619

Caspi, A., Shih, A. Y., Warren, H., Winebarger, A. R., Woods, T. N., Cheung, C. M. M., et al. (2020). *The CubeSat Imaging X-ray Solar Spectrometer (CubIXSS)*. Paper presented at AGU Fall Meeting 2020, Abstract #SH048-0007, held virtually.

Caspi, A., Woods, T. N., & Warren, H. P. (2015). New Observations of the Solar 0.5-5 keV Soft X-Ray Spectrum. *The Astrophysical Journal*, 802(1), L2. https://doi.org/10.1088/2041-8205/802/1/L2

Chartier, A. T., Jackson, D. R., & Mitchell, C. N. (2013). A comparison of the effects of initializing different thermosphere-ionosphere model fields on storm time plasma density forecasts. *Journal of Geophysical Research: Space Physics*, 118(11), 7329–7337. https://doi.org/10.1002/2013JA019034

Cohen, I. J., Anderson, B. J., Bonnell, J. W., Lysak, R. L., Lessard, M. R., Michell, R. G., & Varney, R. H. (2020). Rocket Investigation of Current Closure in the Ionosphere (RICCI): A novel application of CubeSats from a sounding rocket platform. *Advances in Space Research*, 66, 98. https://doi.org/10.1016/j.asr.2019.04.036

DeForest, C. E., de Koning, C. A., & Elliott, H. A. (2017). 3D Polarized Imaging of Coronal Mass Ejections: Chirality of a CME. *The Astrophysical Journal*, 850(2), 130. https://doi.org/10.3847/1538-4357/aa94ca

DeForest, C. E., Howard, T. A., & Tappin, S. J. (2013). The Thomson Surface. II. Polarization. *The Astrophysical Journal*, 765(1), 44. https://doi.org/10.1088/0004-637X/765/1/44

DeForest, C. E., Howard, T. A., Webb, D. F., & Davies, J. A. (2016). The utility of polarized heliospheric imaging for space weather monitoring. *Space Weather*, 14(1), 32–49. https://doi.org/10.1002/2015SW001286

Diard, T., de la Barrière, F., Ferrec, Y., Guérineau, N., Rommeluère, S., Le Coarer, E., & Martin, G. (2016). Compact high-resolution micro-spectrometer on chip: spectral calibration and first spectrum. *Micro- and Nanotechnology Sensors, Systems, and Applications VIII*, *Proc. SPIE*, 9836, 98362W. https://doi.org/10.1117/12.2223692

Dunlop, M. W., Balogh, A., Glassmeier, K.-H., & Robert, P. (2002). Four-point Cluster application of magnetic field analysis tools: The Curlometer. *Journal of Geophysical Research: Space Physics*, 107(A11), 1384. https://doi.org/10.1029/2001JA005088

Eastes, R. W., McClintock, W. E., Burns, A. G., Anderson, D. N., Andersson, L., Codrescu, M., et al. (2017). The Global-Scale Observations of the Limb and Disk (GOLD) Mission. *Space Science Reviews*, 212(1–2), 383–408. https://doi.org/10.1007/s11214-017-0392-2

Gibson, S. E., Vourlidas, A., Hassler, D. M., Rachmeler, L. A., Thompson, M. J., Newmark, J., et al. (2018). Solar Physics from Unconventional Viewpoints. *Frontiers in Astronomy and Space Sciences*, 5, 32. https://doi.org/10.3389/fspas.2018.00032

Howard, T. A., Tappin, S. J., Odstrcil, D., & DeForest, C. E. (2013). The Thomson Surface. III. Tracking Features in 3D. *The Astrophysical Journal*, 765(1), 45. https://doi.org/10.1088/0004-637X/765/1/45

Immel, T. J., England, S. L., Mende, S. B., Heelis, R. A., Englert, C. R., Edelstein, J., et al. (2018). The Ionospheric Connection Explorer Mission: Mission Goals and Design. *Space Science Reviews*, 214(1), 13. https://doi.org/10.1007/s11214-017-0449-2

Knipp, D. J., & Gannon, J. L. (2019). The 2019 National Space Weather Strategy and Action Plan and Beyond. *Space Weather*, 17(6), 794–795. https://doi.org/10.1029/2019SW002254

Korendyke, C. M., Chua, D. H., Howard, R. A., Plunkett, S. P., Socker, D. G., Thernisien, A. F. R., et al. (2015). MiniCOR: A miniature coronagraph for interplanetary cubesat. *Proceedings of the AIAA/USU Conference on Small Satellites*, Science/Mission Payloads, SSC15-XII-6. http://digitalcommons.usu.edu/smallsat/2015/all2015/82

Lario, D., Aran, A., Gómez-Herrero, R., Dresing, N., Heber, B., Ho, G. C., et al. (2013). Longitudinal and Radial Dependence of Solar Energetic Particle Peak Intensities: STEREO, ACE, SOHO, GOES, and MESSENGER Observations. *The Astrophysical Journal*, 767(1), 41. https://doi.org/10.1088/0004-637X/767/1/41

Liewer, P. C. Klesh, A. T., Lo, M. W., Murphy, N., Staehle, R. L., Angelopoulos, A., et al. (2014). A Fractionated Space Weather Base at $L_5$ using CubeSats and Solar Sails. In M. Macdonald (Ed.), *Advances in Solar Sailing* (pp. 269–288). Chichester, UK: Springer Praxis. https://doi.org/10.1007/978-3-642-34907-2_19

Lo, M. W., Llanos, P. J., & Hintz, G. R. (2010). An $L_5$ Mission to Observe the Sun and Space Weather, Part I. *Proceedings of the AAS/AIAA Space Flight Mechanics Meeting*, AAS 10-121. https://dx.doi.org/10.13140/2.1.5166.0162

Lugaz, N., Farrugia, C. J., Winslow, R. M., Al-Haddad, N., Galvin, A. B., Nieves-Chinchilla, T., et al. (2018). On the Spatial Coherence of Magnetic Ejecta: Measurements of Coronal Mass Ejections by Multiple Spacecraft Longitudinally Separated by 0.01 AU. *The Astrophysical Journal*, 864(1), L7. https://doi.org/10.3847/2041-8213/aad9f4

Marshall, R. A., Xu, W., Woods, T., Cully, C., Jaynes, A., Randall, C., et al. (2020). The AEPEX mission: Imaging energetic particle precipitation in the atmosphere through its bremsstrahlung X-ray signatures. *Advances in Space Research*, 66(1), 66–82. https://doi.org/10.1016/j.asr.2020.03.003

Martineau, R. J., Pratt, J., & Swenson, C. (2015). *The Auroral Spatial Structures Probe: magnetic and electric field measurements during an active aurora at fine spatial and temporal scales*. Paper presented at AGU Fall Meeting 2015, Abstract #SA31F-2384, San Francisco, CA.

Mason, J. P., Baumgart, M., Rogler, B., Downs, C., Williams, M., Woods, T. N., et al. (2017). MinXSS-1 CubeSat On-Orbit Pointing and Power Performance: The First Flight of the Blue Canyon Technologies XACT 3-axis Attitude Determination and Control System. *Journal of Small Satellites*, 6(3), 651–662.

Mason, J. P., Chamberlin, P. C., Seaton, D., Burkepile, J., Colaninno, R., Dissauer, K., et al. (2021). SunCET: The Sun Coronal Ejection Tracker Concept. *Journal of Space Weather and Space Climate*, 11, 20. https://doi.org/10.1051/swsc/2021004

Mason, J. P., Woods, T. N., Caspi, A., Chamberlin, P. C., Moore, C., Jones, A., et al. (2016). Miniature X-Ray Solar Spectrometer: A Science-Oriented, University 3U CubeSat. *Journal of Spacecraft and Rockets*, 53(2), 328–339. https://doi.org/10.2514/1.A33351

Mason, J. P., Woods, T. N., Chamberlin, P. C., Jones, A., Kohnert, R., Schwab, B., et al. (2020). MinXSS-2 CubeSat mission overview: Improvements from the successful MinXSS-1 mission. *Advances in Space Research*, 66(1), 3–9. https://doi.org/10.1016/j.asr.2019.02.011

Masutti, D., Denis, A. Wicks, R., Thoemel, J., Kataria, D., Smith, A., & Muylaert, J. (2018). *The QB50 Mission for the Investigation of the Mid-lower Thermosphere: Preliminary Results and Lessons Learned*. Paper presented at 15th Annual International Planetary Probe Workshop (IPPW-2018) Short Course, Boulder, CO.

Millan, R. M., Sotirelis, T., Sample, J. G., Woodger, L. A., Griffith, B. A. A., Nikoukar, R., et al. (2020). *REAL: A CubeSat Mission to Study Energetic Electron Precipitation into Earth's Atmosphere*. Paper presented at AGU Fall Meeting 2020, Abstract #SM009-11.

Mooney, M. K., Marsh, M. S., Forsyth, C., Sharpe, M., Hughes, T., Bingham, S., et al. (2021). Evaluating Auroral Forecasts Against Satellite Observations. *Space Weather*, 19(8), e02688. https://doi.org/10.1029/2020SW002688

Moore, C. S., Caspi, A., Woods, T. N., Chamberlin, P. C., Dennis, B. R., Jones, A. R., et al. (2018). The Instruments and Capabilities of the Miniature X-Ray Solar Spectrometer (MinXSS) CubeSats. *Solar Physics*, 293(2), 21. https://doi.org/10.1007/s11207-018-1243-3

Moretto, T., & Robinson, R. M. (2008). Small Satellites for Space Weather Research. *Space Weather*, 6(5), 05007. https://doi.org/10.1029/2008SW000392

National Research Council (2013). *Solar and Space Physics: A Science for a Technological Society*. Washington, DC: The National Academies Press. https://doi.org/10.17226/13060

National Academies of Sciences, Engineering, and Medicine (2016). *Achieving science with CubeSats: Thinking inside the box*. Washington, DC: The National Academies Press. https://doi.org/10.17226/23503

National Science and Technology Council (2015a). *National Space Weather Strategy*. Washington, DC: Executive Office of the President (EOP). Retrieved from https://www.sworm.gov/publications/2015/nsws_final_20151028.pdf

National Science and Technology Council (2015b). *National Space Weather Action Plan*. Washington, DC: Executive Office of the President (EOP). Retrieved from https://www.sworm.gov/publications/2015/swap_final__20151028.pdf

National Science and Technology Council (2019). *National Space Weather Strategy and Action Plan*. Washington, DC: Executive Office of the President (EOP). Retrieved from https://www.whitehouse.gov/wp-content/uploads/2019/03/National-Space-Weather-Strategy-and-Action-Plan-2019.pdf

Nieves-Chinchilla, T., Lal, B., Robinson, R., Caspi, A., Jackson, D. R., Moretto Jørgensen, T., & Spann, J. (2020). International Coordination and Support for SmallSat-Enabled Space Weather Activities. *Space Weather*, 18(12), e02568. https://doi.org/10.1029/2020SW002568

Rodgers, E. M., Bailey, S. M., Warren, H. P., Woods, T. N., & Eparvier, F. G. (2006). Soft X-ray irradiances during solar flares observed by TIMED-SEE. *Journal of Geophysical Research: Space Physics*, 111(A10), A10S13. https://doi.org/10.1029/2005JA011505

Ruf, C., Unwin, M., Dickinson, J., Rose, R., Rose, D., Vincent, M., & Lyons, A. (2013). CYGNSS: Enabling the Future of Hurricane Prediction [Remote Sensing Satellites]. *IEEE Geoscience and Remote Sensing Magazine*, 1(2), 52–67. https://doi.org/10.1109/MGRS.2013.2260911

Sojka, J. J., Jensen, J., David, M., Schunk, R. W., Woods, T., & Eparvier, F. (2013). Modeling the ionospheric E and F1 regions: Using SDO-EVE observations as the solar irradiance driver. *Journal of Geophysical Research: Space Physics*, 118(8), 5379–5391. https://doi.org/10.1002/jgra.50480

Sojka, J. J., Jensen, J. B., David, M., Schunk, R. W., Woods, T., Eparvier, F., et al. (2014). Ionospheric model-observation comparisons: E layer at Arecibo Incorporation of SDO-EVE solar irradiances. *Journal of Geophysical Research: Space Physics*, 119(5), 3844–3856. https://doi.org/10.1002/2013JA019528

Spence, H. E., Caspi, A., Bahcivan, H., Nieves-Chinchilla, J., Crowley, G., Cutler, J., et al. (2022), Recent Achievements and Lessons Learned from Small Satellite Missions for Space Weather-Oriented Research. *Space Weather*, submitted (this issue). https://doi.org/10.1029/2021SW003031

Thoemel, J., Singarayar, F., Scholz, T., Masutti, D., Testani, P., Asma, C., et al. (2014). Status of the QB50 CubeSat Constellation Mission. *Proceedings of the 65th International Astronautical Congress (IAC-2014)*, IAC-14.B4.2.11. Retrieved from https://www.qb50.eu/index.php/tech-docs/category/26-ref.html

Verkhoglyadova, O. P., Bussy-Virat, C. D., Caspi, A., Jackson, D. R., Kalegaev, V., Klenzing, J., et al. (2021). Addressing Gaps in Space Weather Operations and Understanding With Small Satellites. *Space Weather*, 19(3), e02566. https://doi.org/10.1029/2020SW002566

Vourlidas, A. (2015). Mission to the Sun-Earth L$_5$ Lagrangian Point: An Optimal Platform for Space Weather Research. *Space Weather*, 13(4), 197–201. https://doi.org/10.1002/2015SW001173

Woods, T. N., Caspi, A., Chamberlin, P. C., Jones, A., Kohnert, R., Mason, J. P., et al. (2017). New Solar Irradiance Measurements from the Miniature X-Ray Solar Spectrometer CubeSat. *The Astrophysical Journal*, 835(2), 122. https://doi.org/10.3847/1538-4357/835/2/122

Zheng, Y., Lynch, K. A., Boehm, M., Goldstein, R., Javadi, H., Schuck, P., et al. (2003). Multipoint measurements of field-aligned current density in the auroral zone. *Journal of Geophysical Research (Space Physics)*, 108(A5), 1217. https://doi.org/10.1029/2002JA009450